\documentclass[aps,prd,preprintnumbers,showpacs,showkeys,nofootinbib,
superscriptaddress,fleqn,floatfix,tightenlines,10pt]{revtex4}
\usepackage{amsmath,amsfonts,amssymb,amscd,amsxtra,amsthm}
\usepackage{graphicx}  
\usepackage{epstopdf}
\usepackage{dcolumn}  
\usepackage{bm}          
\usepackage{slashed}
\usepackage{float}
\usepackage{mathtools}
\usepackage{amsbsy}
\usepackage{amstext}

\usepackage[utf8]{inputenc} 
\usepackage{booktabs}
\usepackage[normalem]{ulem} 
\usepackage[dvipsnames]{xcolor} 
\usepackage{tabularx}
\usepackage{array} 
\usepackage{slashed}
\usepackage{tikz}
\renewcommand\sout{\bgroup \color{red} \ULdepth=-.5ex \ULset}

\makeatletter

\begin{document}  
\preprint{INHA-NTG-04/2019}
\title{Electromagnetic properties of singly heavy baryons}
\author{June-Young Kim}
\email[E-mail: ]{junyoung.kim@inha.edu}
\affiliation{Department of Physics, Inha University, Incheon 22212,
Republic of Korea}

\author{Hyun-Chul Kim}
\email[E-mail: ]{hchkim@inha.ac.kr}
\affiliation{Department of Physics, Inha University, Incheon 22212,
Republic of Korea}
\affiliation{School of Physics, Korea Institute for Advanced Study 
  (KIAS), Seoul 02455, Republic of Korea}
\affiliation{  Advanced Science Research Center, Japan Atomic 
Energy Agency, Shirakata, Tokai, Ibaraki, 319-1195, Japan}

\date{\today}
\begin{abstract}
In the present talk, we report recent results on electromagnetic
  properties of the baryon decuplet within the SU(3) chiral
  quark-soliton model, taking into account the effects of explicit
  flavor SU(3) symmetry breaking. We focus on the comparison of the
  present results on the electromagnetic form factors of the
  $\Delta^{+}$  baryon with the lattice data.  
\end{abstract}
\maketitle

\section{Introduction}
When we investigate the structure of a baryon, its electromagnetic
(EM) properties are the first subject to study, since it reveals how
the quarks are distributed inside it. While the EM properties of the
baryon octet have been extensively studied both experimentally and
theoretically, those of the baryon decuplet have been examined mainly
theoretically. Except for the $\Omega^-$ baryon, all the members of
the baryon decuplet decay strongly, so that it is extremely difficult
to measure the EM form factors of the baryon decuplet. Nevertheless,
using the electron and photon beams, experimentalists put a great deal
of efforts on extracting information on the EM properties of the
$\Delta$ isobar~\cite{Kotulla:2002cg, Sparveris:2013ena}. On the other
hand, recent investigations in the lattice QCD produce more and more  
quantitative results on the structure of the baryon
decuplet~\cite{Alexandrou:2008bn, Alexandrou:2010jv}.

In the present talk, we will report recent results on the
electromagnetic form factors of the baryon decuplet ~\cite{JYKim},
derived from the self-consistent chiral quark-soliton model
($\chi$QSM)~\cite{Diakonov:1987ty, Blotz:1992pw, Christov:1995vm}.  
The $\chi$QSM provides a simple but useful theoretical framework for
investigating the structure of the baryon decuplet, since the decuplet
representation naturally appears from the  constraint imposed by the
valence quarks. In fact, the EM form factors of the $\Delta$  were already studied within the $\chi$QSM~\cite{Ledwig:2008es} with exact
flavor SU(3) symmetry considered. In the present talk, we 
focus on the effects from explicit breaking of flavor SU(3)
symmetry. The results of the $\Delta$ EM form factors will be compared
with the recent lattice data.

\section{Result and Discussion}
The detailed formalism for deriving the EM form factors of the baryon
decuplet within the $\chi$QSM will appear elsewhere~\cite{JYKim}. In
the present talk, we will concentrate on the results from the model.

\begin{figure}
\centering
\includegraphics[scale=0.3]{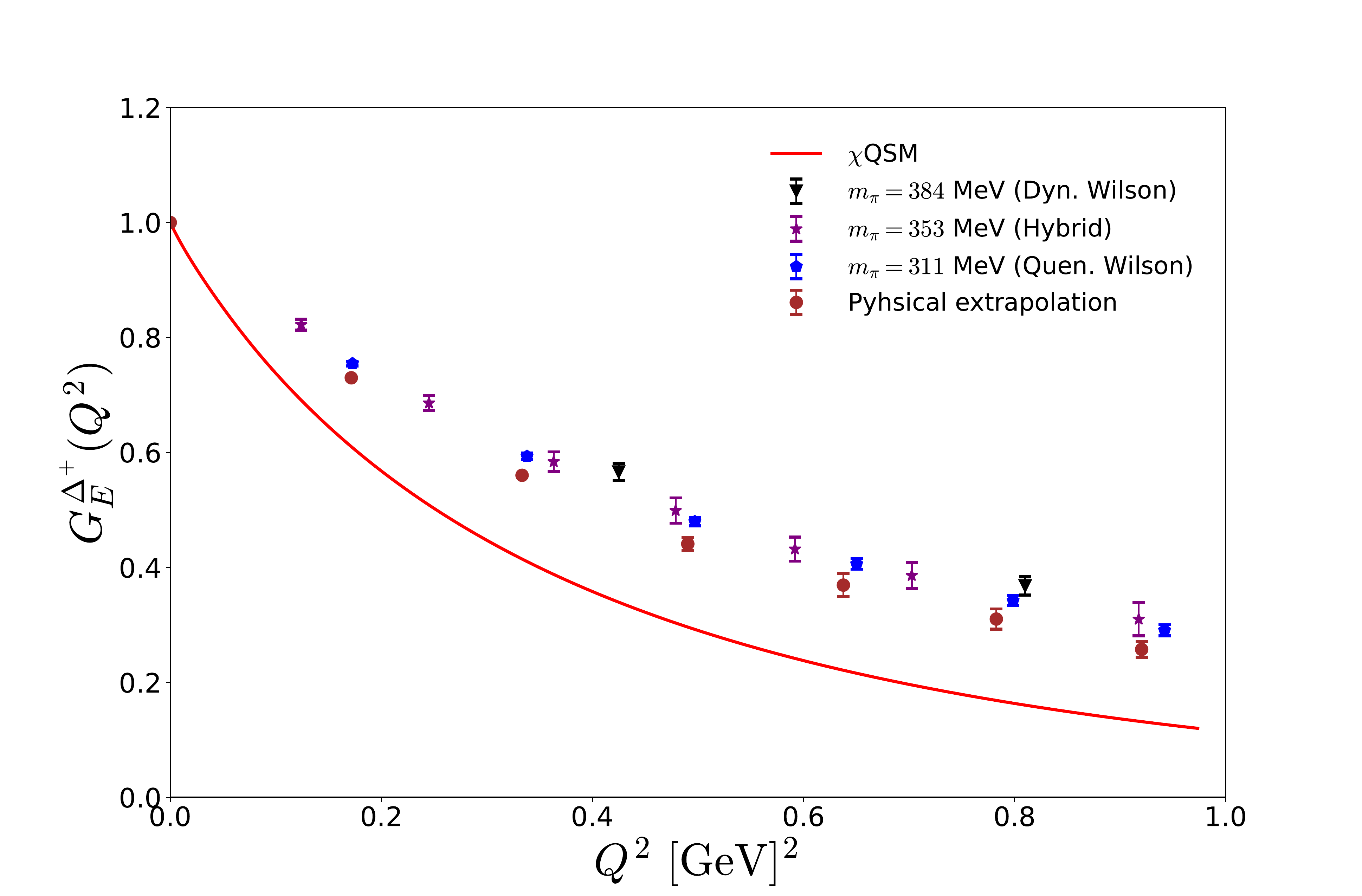}
\caption{The electric monopole form factor of the $\Delta^{+}$ as a
  function of $Q^2$ in comparison with the lattice data. The red curve
  depicts the present result whereas all the points with error bars
  are taken from the lattice QCD
  data~\cite{Alexandrou:2007we,Alexandrou:2008bn}.}  
\label{fig:1}
\end{figure}
In Fig.~\ref{fig:1}, we show the numerical result of the electric
monopole form factor of the $\Delta^{+}$ baryon as a function of $Q^2$
in comparison with the 
lattice data~\cite{Alexandrou:2007we,Alexandrou:2008bn}.
The positive-definite $Q^2$ is defined as $Q^2=-q^2$ in which $q^2$ is
the square of the momentum transfer.  
It is rather well known that when higher unphysical pion masses are
employed the lattice calculation of the nucleon EM form factor yield
the results of hadronic form factors, which fall off more slowly,
compared with the  experimental data. Keeping in mind this, we find
that indeed the present result decreases more rapidly than the lattice
data, as shown clearly in Fig.~\ref{fig:1}. What is interesting is
that the present result even falls off faster than the lattice data
with physical extrapolation. 

\begin{figure}
\centering
\includegraphics[scale=0.3]{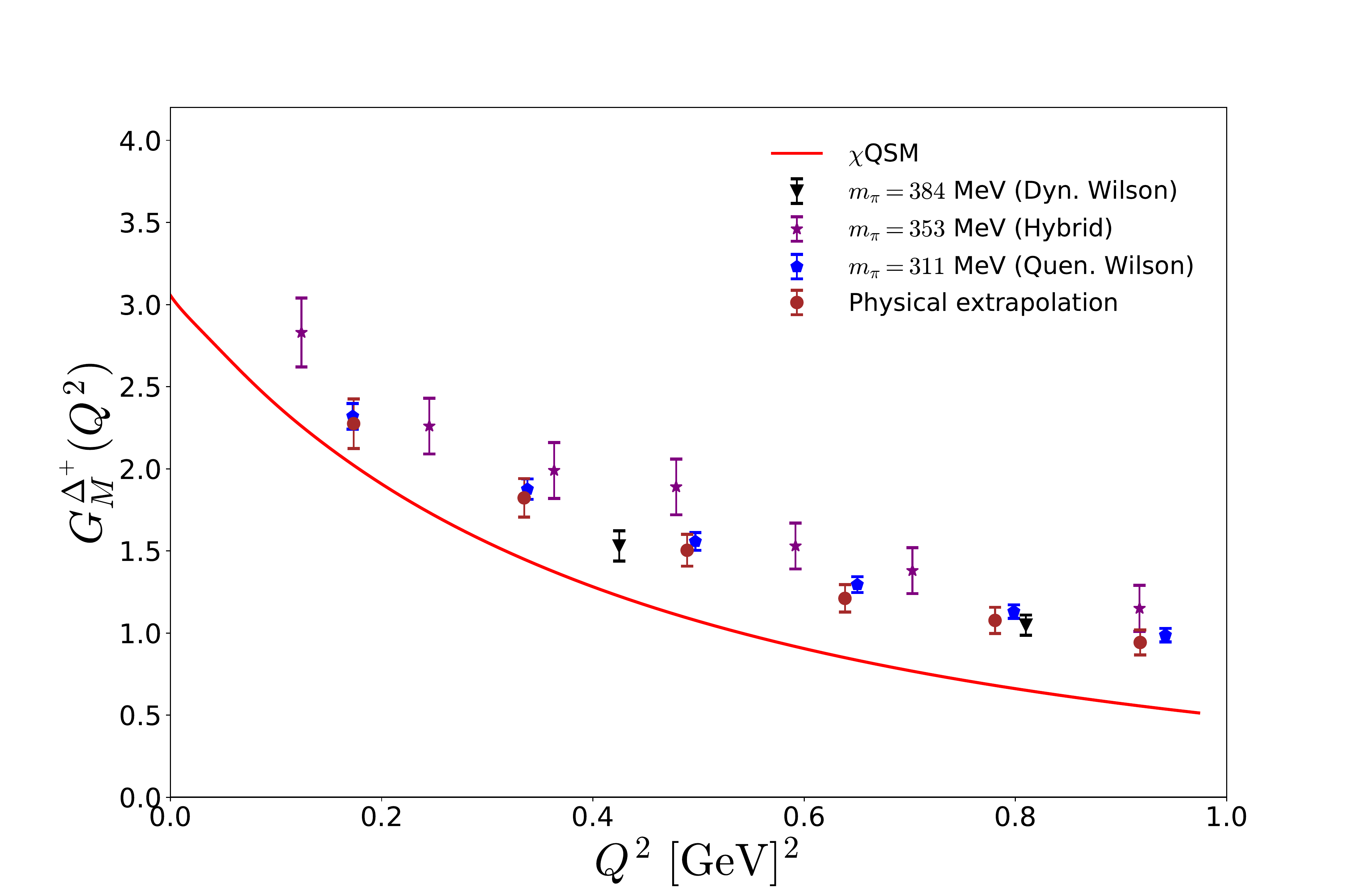}
\caption{The magnetic dipole form factor of the $\Delta^{+}$ as a
  function of $Q^2$ in comparison with the lattice data. Notations are
  the same as in Fig.~\ref{fig:1}.}
\label{fig:2}
\end{figure}
In Fig.~\ref{fig:2}, the numerical result of the $\Delta$ magnetic
dipole form factor is drawn, compared with the lattice data. The
result exhibits a similar tendency as shown in the case of the
electric form factor presented in Fig.~\ref{fig:1}. Again, the reason
can be found in the large value of the pion mass adopted by the lattice
QCD. 

\begin{figure}
\centering
\includegraphics[scale=0.3]{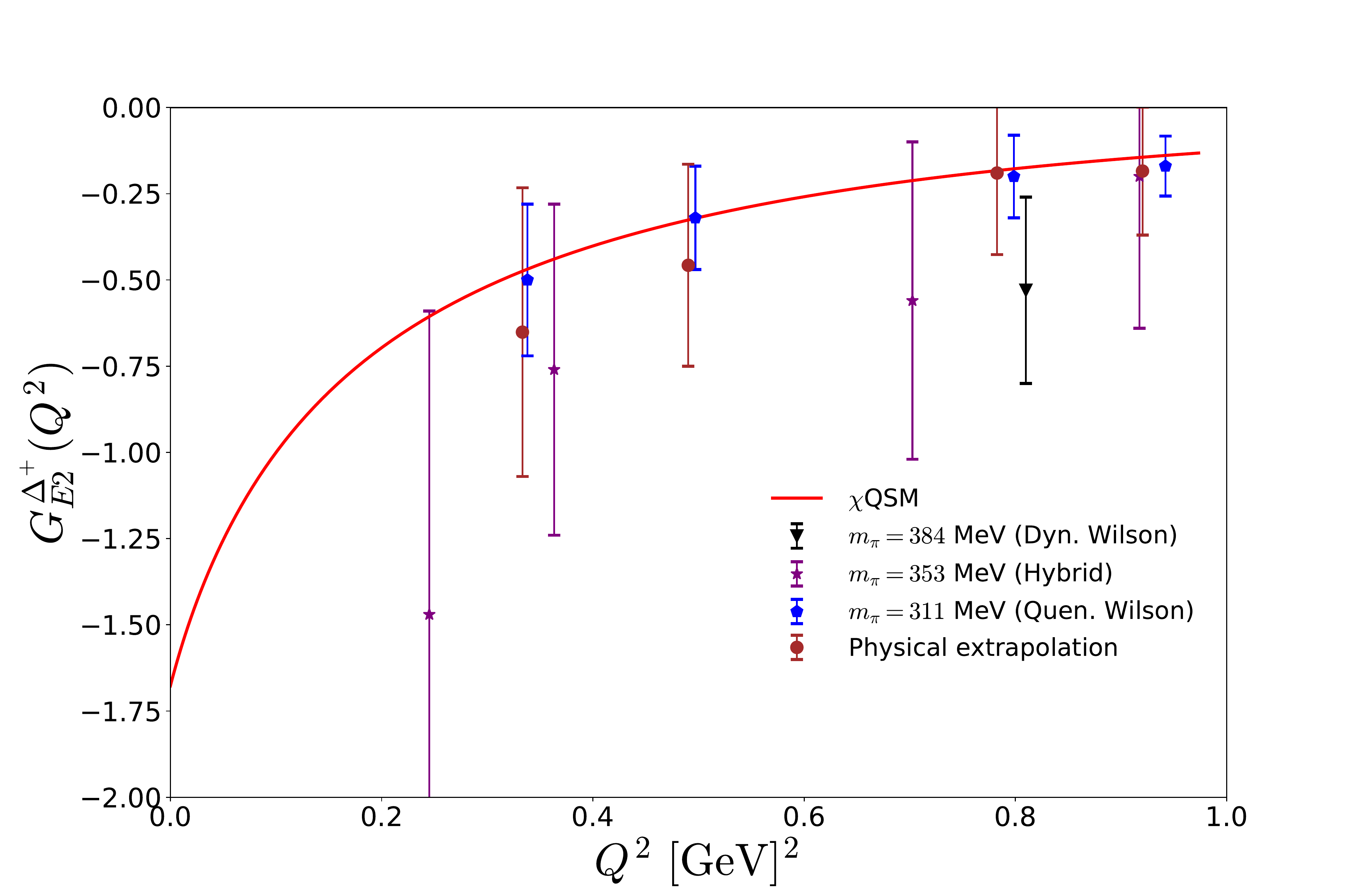}
\caption{The electric quadrupole form factor of the $\Delta^{+}$ as a
  function of $Q^2$ in comparison with the lattice data. Notations are
  the same as in Fig.~\ref{fig:1}.}
\label{fig:3}
\end{figure}
Figure~\ref{fig:3} depicts the result of the electric quadrupole form
factor of the $\Delta^{+}$, again compared  with  those of the
lattice QCD. Unfortunately, the lattice results show large
uncertainty, so that one can not draw any meaning conclusion from this
comparison.  However, the general dependence on $Q^2$ of the present
result seems similar to the  lattice one. 

\begin{table}[tbh]
\centering
\caption{Electromagnetic properties of $\Delta^{+}$ baryon.}
\label{t1}
\begin{tabular}{llll}
\hline
\hline
 & $\langle r^{2} \rangle_{E}^{\Delta^{+}}$$[\mathrm{fm}^{2}]$ &
 $\mu_{\Delta^{+}}$[$\mu_{N}$] & $Q_{\Delta^{+}}[e \cdot \mathrm{fm}^{2}]$ \\ 
 \hline
$\chi$QSM ($m_{s}=180$) & 0.79 & 2.33 & 0.043 \\
$\chi$QSM ($m_{s}=0$) & 0.83 & 2.47 & 0.053 \\
\hline
Exp.~\cite{Kotulla:2002cg} & - & $2.7^{+1.0}_{-1.3}\pm1.5 \pm3$ & - \\
LQCD~\cite{Boinepalli:2009sq} & 0.41(6) & 1.6(3) & - \\
\hline
\hline
\end{tabular}
\end{table}

In Tab.~\ref{t1}, we list the numerical results on the squared
electric charge radius, the magnetic dipole moment, and the electric
quadrupole moment of the $\Delta^+$ with the SU(3) symmetry breaking
taken into account. The inclusion of the SU(3) symmetry-breaking
effects lessens slightly the squared electric charge radius. It
indicates that the SU(3) breaking makes the $\Delta^+$ a little bit
more compact. As already expected from Fig.~\ref{fig:1}, the lattice
data produces a smaller value of $\langle r^2\rangle_E$.
The contribution from the flavor SU(3) symmetry breaking also reduces
the value of the $\Delta^+$ magnetic dipole moment, which is opposite
the case of the nucleon~\cite{Kim:1995mr}. Note that the
nucleon mass for the nuclear magneton is replaced by the classical
soliton mass in the present model in order to keep a consistency, as
already pointed out by Ledwig et al.~\cite{Ledwig:2008es}.
When it comes to the electric quadrupole moment, the effects of the
explicit flavor SU(3) symmetry breaking turns out to be much large in
contrast to the previous observables such as the electric charge
radius and magnetic dipole moment. The reason will be discussed
elsewhere in detail~\cite{JYKim}.

We want to mention that the magnetic quadrupole form factor
vanishes within the present framework. This is already well known that
in any chiral model with hedgehog symmetry the magnetic octupole form
factor  becomes  null. However, it indicates that the magnetic octupole form
factor will be negligibly small in nature.  Finally, we find that
$U$-spin symmetry is preserved in the whole form factors of the baryon 
decuplet in the flavor SU(3) symmetric case.

\section{Summary and conclusion}
In this talk, we presented recent results of the electromagnetic form
factors of the $\Delta^+$ isobar in comparison with the lattice QCD
data, and discuss the related observables with the effects of flavor
SU(3) symmetry breaking. While the present talk concentrates on the
electromagnetic properties of the $\Delta^+$ isobar, we can easily
extend the present work to examine those of all other members
belonging to the baryon decuplet. A more complete article will soon
appear~\cite{JYKim}. 

\section*{Acknowledgment} 
The work is supported by Basic Science Research Program through the
National Research Foundation (NRF) of Korea funded by the 
Korean government (Ministry of Education, Science and
Technology(MEST)): Grant No. NRF-2018R1A2B2001752.

\end{document}